\documentstyle[mprocl,psfig]{article}

\def\be{\begin{equation}}
\def\ee{\end{equation}}
\def\bea{\begin{eqnarray}}
\def\eea{\end{eqnarray}}

\begin{document}

\title{QUASICLASSICAL CALCULATION OF SPONTANEOUS CURRENT IN
RESTRICTED GEOMETRIES}

\author{M.H.S. Amin, M. Coury, A.M. Zagoskin\footnote{Also at Physics
and Astronomy Dept., The University of British Columbia, 6224
Agricultural Rd., Vancouver, B.C., V6T 1Z1, Canada.}}

\address{D-Wave Systems Inc., 320-1985 W.~Broadway, Vancouver, B.C.,
V6J 4Y3, Canada}

\author{S.N. Rashkeev}

\address{Dept.~of Physics and Astronomy, Vanderbilt
University, Box 1807 Station B,\\ Nashville, TN 37235, USA}

\author{A.N. Omelyanchouk}

\address{B.I. Verkin Institute for Low Temperature Physics and
Engineering,\\
Ukrainian National Academy of Sciences, Lenin Ave.~47, Kharkov
310164, Ukraine}


\maketitle\abstracts{ Calculation of current and order parameter
distribution in inhomogeneous superconductors is often based on a
self-consistent solution of Eilenberger equations for
quasiclassical Green's functions. Compared to the original Gorkov
equations, the problem is much simplified due to the fact that
the values of Green's functions at a given point are connected to
the bulk ones at infinity (boundary values) by ``dragging'' along
the classical trajectories of quasiparticles. In finite size
systems, where classical trajectories undergo multiple
reflections from surfaces and interfaces, the usefulness of the
approach is no longer obvious, since there is no simple criterion
to determine what boundary value a trajectory corresponds to, and
whether it reaches infinity at all. Here, we demonstrate the
modification of the approach based on the Schophol -Maki
transformation, which provides the basis for stable numerical
calculations in 2D. We apply it to two examples: generation of
spontaneous currents and magnetic moments in isolated islands of
{\it d}-wave superconductor with subdominant order-parameters $s$
and $d_{xy}$, and in a grain boundary junction between two
arbitrarily oriented {\it d}-wave superconductors. Both examples
are relevant to the discussion of time-reversal symmetry breaking
in unconventional superconductors, as well as for application in
quantum computing.}

\section{Introduction}

Pairing symmetry of unconventional superconductors can produce
time-reversal symmetry breaking states.~\cite{review}  Especially
interesting are high T$_{c}$ cuprates, with their $d$-wave
symmetry, since the recently developed
technology~\cite{tzalenchuk} allows fabrication of such
structures, with controllable characteristics, as
$\pi$-junctions, submicron size
``$\phi_{0}$"-junctions~\cite{ilichev} (i.e. junctions with
equilibrium phase difference $\phi_0$ which is neither 0 nor
$\pi$),~\cite{amin1} $\pi$-SQUIDs~\cite{schulz} or
$\pi/2$-SQUIDs,~\cite{ACR} and superconducting qubit
prototypes.~\cite{qubit} Therefore, quantitative prediction of
properties of such restricted systems becomes relevant not just
from the theoretical point of view.

General approach to such calculations is based on Gorkov
equations for Green's functions of the superconductor. It is
limited only by the applicability of BCS-like ``mean field''
description of the superconducting state. This description is now
considered valid on the phenomenological level,~\cite{book}
independently from further developments in the first principles'
theory of high T$_{c}$ superconductivity. Quasiclassical limit of
these equations, Eilenberger equations,~\cite{Eilenberger} is
strictly speaking valid only when $|\Delta |\ll E_{F},$. Unlike in
conventional superconductors, this condition is not completely
satisfied in high $T_c$ superconductors but is still a good
approximation.

\section{Quasiclassical Approach}

We use the standard approach based on Eilenberger
equations~\cite{Eilenberger} for quasiclassical Green's functions
\begin{equation}
{\bf v}_{F}\cdot \nabla \widehat{g}+[\omega \widehat{\tau }_{3}+
\widehat{\Delta },\widehat{g}]=0, \label{eil}
\end{equation}
with normalization condition $\widehat{g}^{\ 2}=\widehat{1}$,
where $\omega $ is the Matsubara frequency and
\[
\widehat{\tau}_3=\left(
\begin{array}{cc}
1  & 0  \\
0 & -1
\end{array}
\right) ,\quad \widehat{g}=\left(
\begin{array}{cc}
g  & f  \\
f ^{\dagger } & -g
\end{array}
\right) ,\quad \widehat{\Delta }=\left(
\begin{array}{cc}
0 & \Delta \\
\Delta ^{\dagger } & 0
\end{array}
\right) .
\]
The matrix Green's function $\widehat{g}$ and the superconducting
order parameter $\Delta$ are both functions of the Fermi velocity
${\bf v}_{F}$ and position ${\bf r}$. $\Delta$ is determined by
the (2D) self-consistency equation
\begin{eqnarray}
\Delta (\theta)=2\pi N(0)T\sum\limits_{\omega
>0} \left< V_{\theta\theta'}f (\theta')
\right>_{\theta'}  \nonumber
\end{eqnarray}
where $\theta$ is the angle between ${\bf v}_F$ and the $x$-axis,
$V_{\theta\theta'}$ interaction potential, $N(0)$ density of
states at the Fermi surface, and $\left<...\right>_\theta$
represents averaging over $\theta$. Generally, it is possible to
obtain a mixture of different symmetries of the order parameter,
e.g. $\Delta =\Delta _{x^{2}-y^{2}}+\Delta _{xy}+\Delta _{s}$
where $\Delta _{x^{2}-y^{2}}=\Delta _{1}\cos 2\theta ,$ $\Delta
_{xy}=\Delta _{2}\sin 2\theta $, and $\Delta _{s}$ are the
dominant $d_{x^{2}-y^{2}}$ component, and the subdominant
$d_{xy}$ and the $s$ components of the order parameter
respectively. The corresponding interaction potential, $V_{\theta
\theta ^{\prime }}=V_{d1}\cos 2\theta \cos 2\theta ^{\prime
}+V_{d2}\sin 2\theta \sin 2\theta ^{\prime }+V_{s},$ must be
substituted in the self-consistency equation. The current density
${\bf j(r)}$ is found from $g$ as
\begin{eqnarray}
{\bf j}=-4\pi ieN(0)T\sum\limits_{\omega >0} \left< {\bf v}_{F}g
\right>_{\theta } \nonumber
\end{eqnarray}

For numerical calculations, it is
conventional~\cite{amin1,amin2,Schopohl} to parameterize the
quasiclassical Green's functions by so called coherent functions
$a,b$ via

\begin{equation}
g={\frac{1-ab}{1+ab}}\ ,\quad f={\frac{2a}{1+ab}}. \label{EqC1}
\end{equation}
Functions $a$ and $b$ satisfy two independent, but nonlinear,
equations
\begin{eqnarray}
{\bf v}_{F}\cdot \nabla a &=&\Delta -\Delta ^{\ast }a^{2}-2\omega
a
\nonumber \\
-{\bf v}_{F}\cdot \nabla b &=&\Delta ^{\ast }-\Delta
b^{2}-2\omega b. \label{EqC2}
\end{eqnarray}
From these equations it follows that $a(-{\bf v}_{F})=b^{\ast
}({\bf v}_{F})$ and $b(-{\bf v}_{F})=a^{\ast }({\bf v}_{F})$. One
should solve these equations along all possible quasiclassical
trajectories and perform the summation over the trajectories to
calculate the order parameter current density. To find $a$ and $b$
along the trajectories, one needs to use boundary conditions at
the ends of the trajectories. In macroscopically large systems,
one usually assumes that {\em all} the trajectories go deep into
the bulk of the superconductor, i.e., it is possible to use the
bulk solutions
\begin{equation}
a_{\pm }={\frac{\Delta }{\omega \pm \Omega }}\ ,\qquad b_{\pm
}={\frac{\Delta ^{\ast }}{\omega \pm \Omega }}  \label{EqC3}
\end{equation}
with $\Omega =\sqrt{\omega ^{2}+|\Delta |^{2}}$, as the boundary
conditions at infinity. In the case of ``restricted'' systems, the
above assumption is no longer self-evident. Nevertheless, we will
see that a stable numerical procedure can be still developed.

Numerical calculation is stable if the integration for $a$ ($b$)
is taken in the (opposite) direction of ${\bf v}_{F}$. When
$\Delta $ is a constant, the solution of Eq.~(\ref{EqC2}) for $a$
is
\begin{eqnarray}
a_{f} &=& a_{+}+{\frac{a_{i}-a_{+}}{1+{\frac{\Delta ^{\ast }}{\Omega }}%
(a_{i}-a_{-})e^{\Omega \tau }\sinh \Omega \tau }} \nonumber \\
&\approx& a_{+}+{\frac{\Omega }{\Delta ^{\ast }}}\left(
{\frac{a_{i}-a_{+}}{a_{i}-a_{-}}}\right) e^{-2\Omega \tau },
\quad {\rm for} \ \Omega \tau \gg 1
\end{eqnarray}
where $a_{i}$ and $a_{f}$ are the values of $a$ at the initial
(${\bf r}_{i}$) and final (${\bf r}_{f}$) points of the
trajectory, and $\tau =|{\bf r} _{f}-{\bf r}_{i}|/v_{F}$ is
proportional to the distance between the initial and final points
along the trajectory. It is clear that the solution for $a$
relaxes to the bulk value $a_{+}$ at the distance $L={\bf
v}_{F}/2\Omega $ which is of the order of the coherence length
$\xi _{0}$. In other words, when the quasiparticle moves away from
the initial point at a distance of a few $\xi _{0}$'s, any
information about the initial point $a_{i}$ becomes lost. This
observation is also valid for $b$, and is crucial for what that
follows.

\begin{figure}
\center \hspace{0mm} \psfig{figure=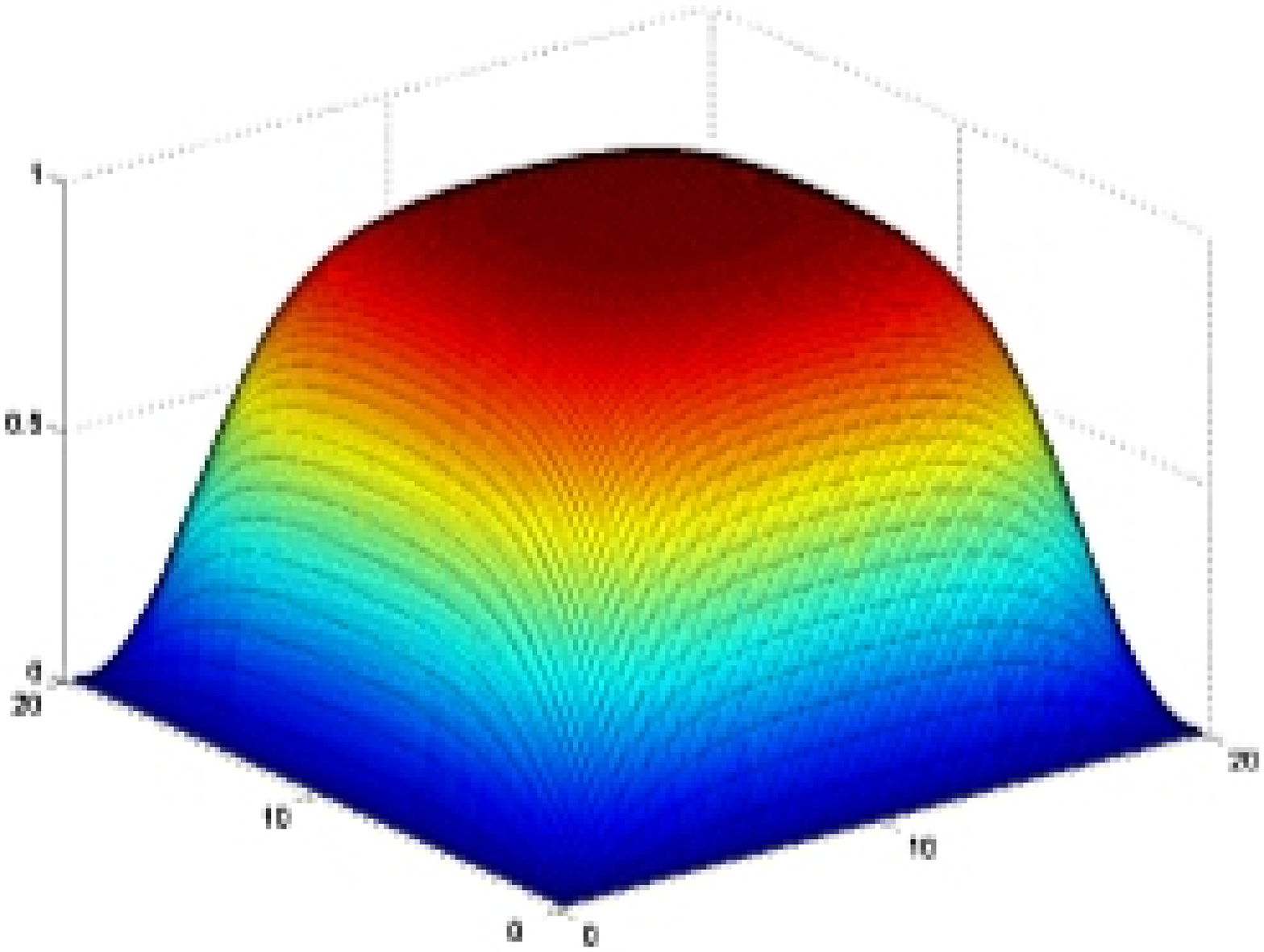,height=1.3in}
\hspace{1cm} \psfig{figure=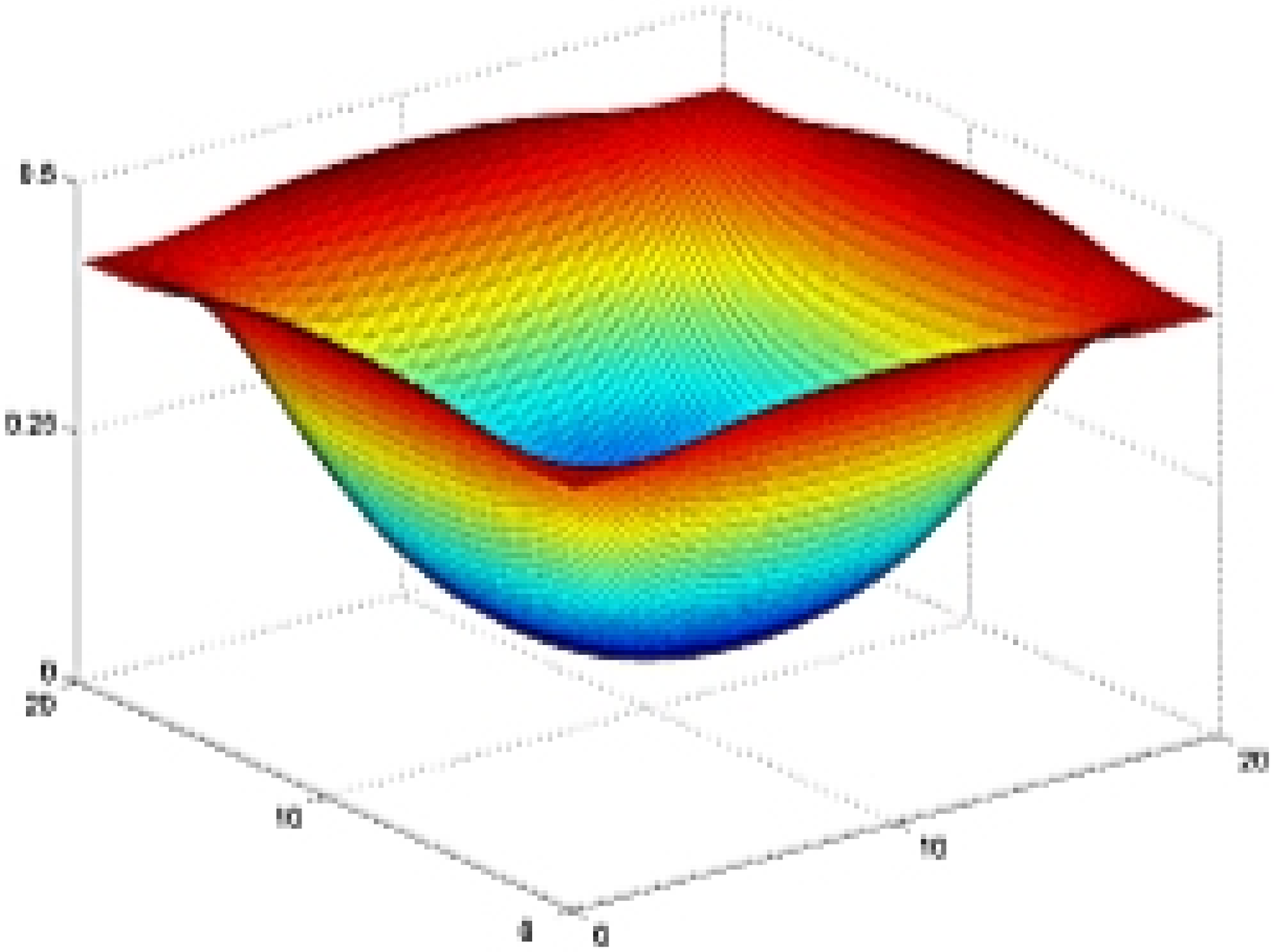,height=1.3in}
\caption{Absolute values of the dominant (left) and subdominant
(right) order parameters for a small square $d$-wave super
conductor in the presence of subdominant $s$-wave order
parameter. The orientation of the main order parameter is
$45^\circ$ rotated with respect to the boundaries.} \label{fig1}
\end{figure}

Let us now specifically consider a restricted system. After
integrating over a few $\xi _{0}$'s, $a_{f}$ will be almost
independent of $a_{i}$, although it may never coincide with the
bulk value $a_{+}$. This solution corresponds to a simple
exponential relaxation of the functions $a$ and $b$ to their
local ``steady-state'' values defined by the local value of the
order parameter. This value is the limit for the functions $a$
and $b$ at this spatial point. Such relaxation of Green's
functions significantly simplifies the numerical solution of the
self-consistent two-dimensional problem. The system therefore has
no memory of the local values of $\Delta $ beyond several $\xi
_{0}$ along the trajectory.

In order to calculate $a$, we define all possible trajectories
going through a given point of the system. Along each of them we
move back (in the direction opposite to ${\bf v}_{F}$) a cutoff
distance (about $10\xi _{0}$-- $20\xi _{0})$ and choose that
point as the beginning of the trajectory. We set the bulk
solution ($a_{+}$) as the initial value for $a$ at that point
(this really does not matter too much because the system has no
memory) and integrate along the trajectory, taking into account
the reflections at the boundaries, until we get back to the
calculation point. Calculation for $b$ is the same, except that
the direction of integration is now opposite to ${\bf v}_{F}$.
These calculations are repeated for all trajectories and all
points of the system (on a mesh). After each iteration, we
calculate the new $\Delta$ and use it for the next iteration
until the self-consistency is achieved. We found this method to
be very stable and independent of the value of the ``cutting''
distance. One should note that the above procedure is not valid
in the presence of magnetic field, because a path dependent phase
will be accumulated to the Green's functions, and the above
mentioned relaxation mechanism along the trajectory does not work
anymore.

\section{Results}

\begin{figure}
\center \hspace{0mm} \psfig{figure=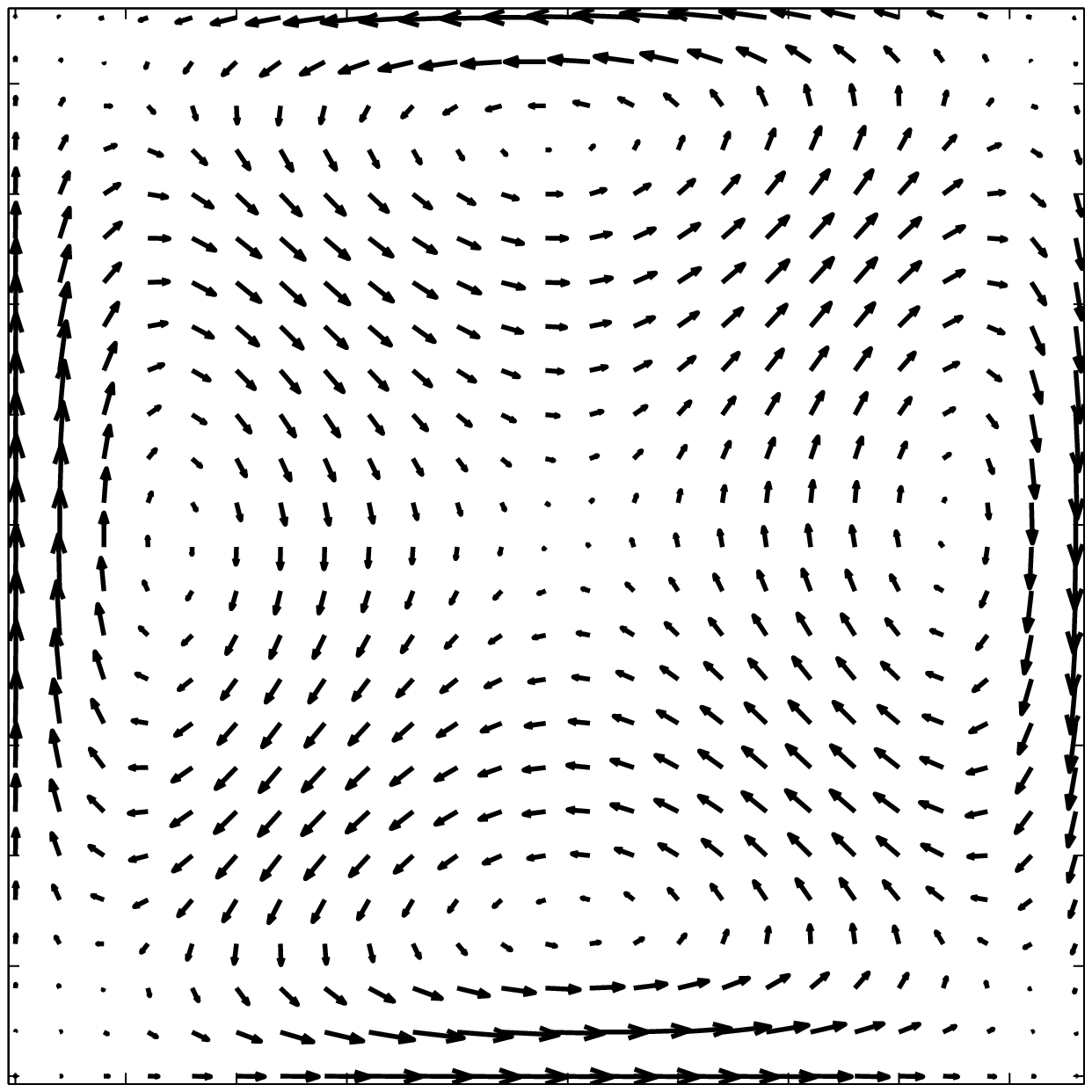,height=1.8in}
\hspace{1cm} \psfig{figure=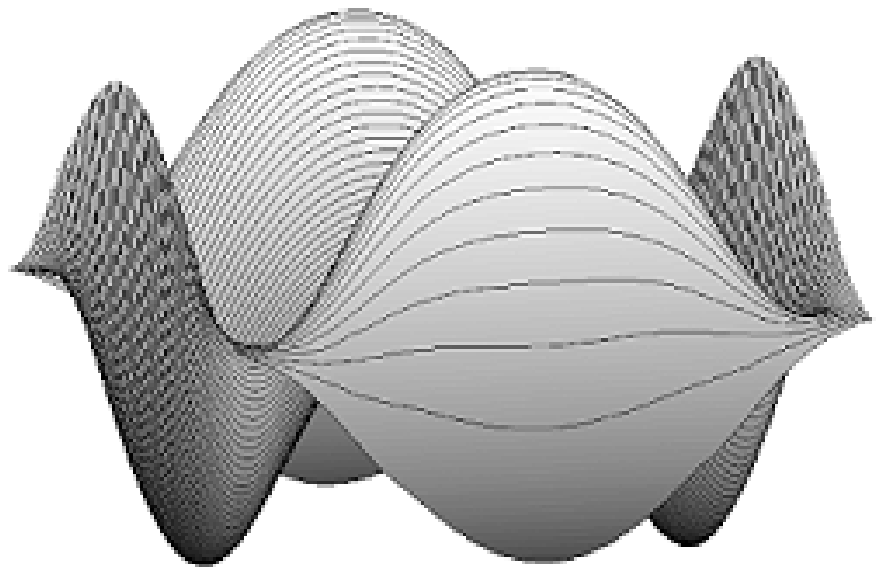,height=1.3in}
\caption{Spontaneous current density (left) and magnetic field
distribution (right) for the system of Fig.~\ref{fig1}.}
\label{fig2}
\end{figure}

To illustrate the approach, we performed self-consistent
calculations of the order parameter in a small ($20\xi_0 \times
20\xi_0$) square region of $d$-wave superconductor in the
presence of a subdominant $s$-wave order parameter. The
crystalographic $a$ and $b$ directions of the dominant order
parameter make $45^\circ$ angle with respect to the boundaries of
the square. We used a random subdominant order parameter at the
first iteration in order to avoid imposing any assumption on the
phase of the second order parameter.

The two components of the order parameter are displayed in
Fig.~\ref {fig1}. The dominant order parameter is suppressed at
the boundaries of the square. This is due to the special
orientation of the order parameter which requires all
quasiparticles to face opposite sign of the order parameter after
reflecting from the boundaries.~\cite{amin1} As a result, the
subdominant order parameter ($s$) with a $\pi/2$ phase shift with
respect to the dominant order parameter, is the main contributor
at the boundaries. The spontaneous current distribution is
displayed in Fig.~\ref{fig2}. The current does not flow in the
same direction at all the four edges, but changes the direction
from one edge to another closing its path towards the center of
the square. The magnetic field produced by the current is also
shown in Fig.~\ref{fig2}. The maximum value of the magnetic field
is of the order of $10^{-3}G$. Notice that the direction of the
filed changes from one edge to another. Thus, the total flux
produced by such a field is zero.

We also have calculated the spontaneous current and magnetic field
distributions in a $d$-wave grain boundary junction. The system
consists of a finite square ($30\xi_0 \times 30\xi_0$) made of
$d$-wave superconductor divided into two equal parts separated by
a grain boundary junction. The order parameter has
$d_{x^{2}-y^{2}}$ symmetry with $0^{\circ }$ orientation on the
left side of the grain boundary and $45^{\circ }$ on the right
side. We include a $d_{xy}$ subdominant order parameter by adding
an attraction potential in that channel (see Ref.~\cite{amin1}).
The magnitude of potential is chosen in such a way to have a
transition temperature $T_{c2}=0.1T_c$ (in the absence of the
dominant order).~\cite{amin1} We choose a phase difference of
$\Delta \phi =\pi /2$ between the two sides. This actually
corresponds to the equilibrium phase difference of the junction
at which the total current passing through the junction is
zero.~\cite{amin1} Calculations are done at $T=0.05T_{c}$.

\begin{figure}[t]
\center \hspace{0mm} 
\psfig{figure=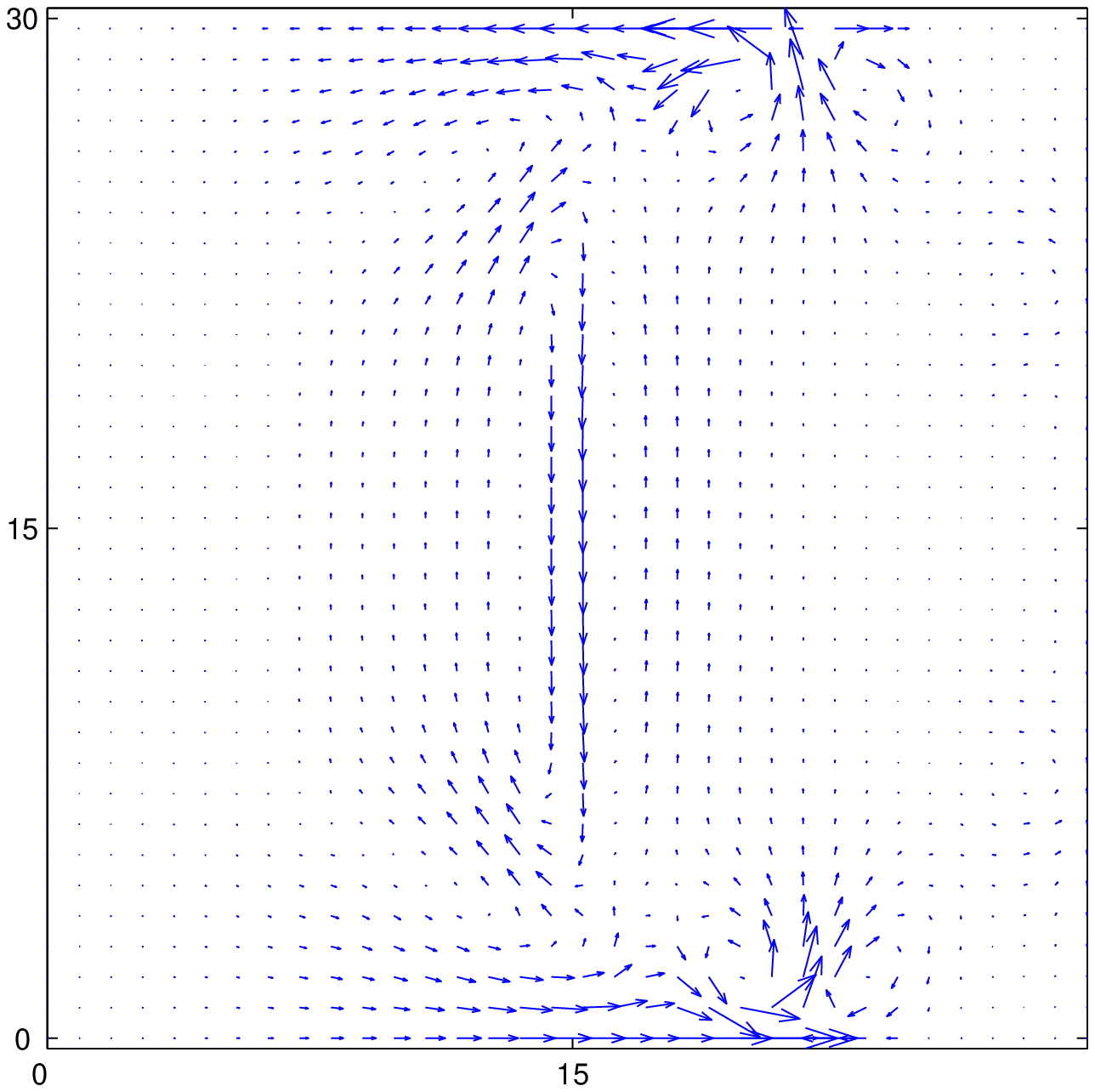,height=1.8in}
\psfig{figure=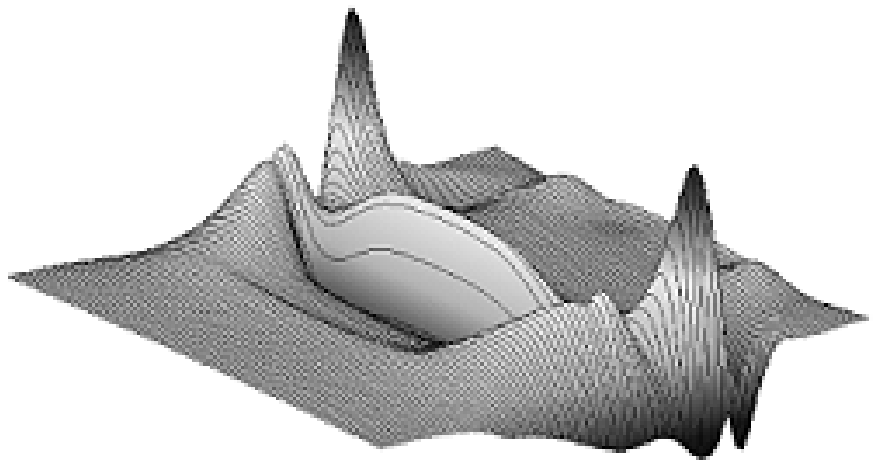,height=1.3in} \caption{Absolute value of
the order parameter for a grain boundary junction between two
d-wave superconductors. The grain boundary is a vertical line
located in the middle. The orientation of the order parameter is
$0^\circ$ on the left and $45^\circ$ on the right.} \label{fig3}
\end{figure}

The results of calculation of the spontaneous current and magnetic
field distributions are displayed in Fig.~\ref{fig3}. Notice that
the current is not symmetric with respect to the grain boundary.
On the left side (with $0^{\circ }$ orientation), the current
returns along the diagonal, whereas on the right side ($45^{\circ
}$ orientation) it forms two vortices and antivortices near the
edges. These vortices are a consequence of the chiral nature of
the $d+id'$ symmetry.~\cite{amin1,amin2} Magnetic field is peaked
at the location of vortices with a maximum of the order of
$10^{-3}G$. It is important to emphasize that, unlike in the
previous case, the existence of the spontaneous current in this
system does not depend on the presence of a subdominant order
parameter (although we assumed a subdominant component here).
Addition of a subdominant order parameter will actually suppress
the spontaneous current at the boundary (see
Refs.~\cite{amin1,amin2}).

\section{Conclusions}

We described a method to calculate equilibrium properties in
finite size superconducting systems. We presented the results of
our calculations for the distribution of spontaneous current and
magnetic field in two systems: a small square region of a
$d$-wave superconductor with a pair breaking boundary, and a
$d$-wave grain boundary junctions between two differently
oriented $d$-wave superconductors. The method described here is
quite general and can be applied to any 2D geometry with proper
boundary conditions. Presence of external magnetic field
invalidates the method described here. The self-generated
magnetic field due to the spontaneous currents, however, is
usually very small so that its effect can be neglected in most
calculations.

\section*{Acknowledgement}

We would like to thank A.~Golubov, A.~Maassen van den Brink,
G.~Rose, and A.Yu. Smirnov for stimulating discussions.

\end{document}